\documentclass[floats,floatfix,showpacs,amssymb,prd,superscriptaddress,twocolumn,aps]{revtex4-1}
\usepackage{graphicx, epsfig, amssymb} 
\usepackage{amsmath, amsfonts}
\usepackage{bm} 

\usepackage[linktocpage]{hyperref}
\usepackage[caption=false]{subfig}
\usepackage[usenames]{color}

\def\be{\begin{equation}}
\def\ee{\end{equation}}
\def\beq{\begin{eqnarray}}
\def\eeq{\end{eqnarray}}

\newcommand{\bea}{\begin{eqnarray}}
\newcommand{\eea}{\end{eqnarray}}
\newcommand{\ben}{\begin{enumerate}}
\newcommand{\een}{\end{enumerate}}
\newcommand{\bi}{\begin{itemize}}
\newcommand{\ei}{\end{itemize}}

\begin{document}

\title{\large Static Einstein-Maxwell black holes with no spatial isometries in $AdS$ space }

 \author{Carlos A. R. Herdeiro}
 \affiliation{Departamento de F\'isica da Universidade de Aveiro and CIDMA, 
 Campus de Santiago, 3810-183 Aveiro, Portugal.}

 \author{Eugen Radu}
 \affiliation{Departamento de F\'isica da Universidade de Aveiro and CIDMA, 
 Campus de Santiago, 3810-183 Aveiro, Portugal.}

\date{June 2016} 

\begin{abstract}
We explicitly construct static black hole solutions to the fully non-linear, $D=4$, Einstein--Maxwell--$AdS$ 
equations that have no continuous spatial symmetries. 
These black holes have a smooth, topologically spherical horizon (section), but without isometries, and approach, asymptotically, global $AdS$ spacetime. 
They are interpreted as bound states of a horizon with the Einstein-Maxwell--$AdS$ solitons 
recently discovered, for appropriate boundary data. 
In sharp contrast with the uniqueness results for Minkowski 
electrovacuum, the existence of these black holes shows that single, equilibrium, BH solutions in $AdS$-electrovacuum admit an arbitrary multipole structure.
\end{abstract}


\maketitle

\date{today}
\noindent{\bf{\em Introduction.}}
In 1967, Israel established a remarkable and influential result in black hole (BH) physics: a static, vacuum, regular (on and outside the horizon) BH in General Relativity (GR) is spherically symmetric~\cite{Israel:1967wq}. A corollary, indeed an application of Birkhoff's theorem, implies that such spacetime is the Schwarzschild BH, hence described by a unique parameter, its ADM mass.  This result, in clear contrast with the \textit{status quo} in other field theories (say, electromagnetism), set the first cornerstone for the celebrated uniqueness theorems~\cite{Chrusciel:2012jk}, establishing the extraordinary simplicity of  BHs in vacuum GR.

Israel's result was swiftly generalized to electrovacuum~\cite{Israel:1967za}, establishing that a \textit{single}, static, BH solution is spherically symmetric and described by only two parameters, its ADM mass and electric charge (excluding magnetic charges).  The purpose of this letter is to establish that the addition of a negative cosmological constant  to the electrovacuum model, hereafter dubbed \textit{$AdS$-electrovacuum}, allows a dramatic departure from Israel's theorem: staticity does not guarantee the existence of \textit{any continuous spatial symmetry}, for physically acceptable BHs.

We establish this result by explicitly constructing the first, fully non-linear,  co-dimension 3, equilibrium, single BH solutions in GR. As examples, we exhibit a sample of exotic BH horizon geometries, deprived of isometries, albeit possessing discrete symmetries, illustrated by their isometric embeddings in Euclidean 3-space.

\noindent{\bf{\em The role of gravitating solitons.}}
%
Gravitating solitons are stationary, everywhere regular spacetimes with localized energy, $i.e.$ \textit{particle-like} solutions of GR (or extensions thereof). Influential examples, focusing on trivial spacetime topologies, have been found, $e.g.$, in Einstein-complex-Klein-Gordon theory, dubbed boson stars~\cite{Kaup:1968zz,Ruffini:1969qy,Schunck:2003kk}, or in Einstein-Yang-Mills (EYM) theory
\cite{Bartnik:1988am}. 
When gravitating solitons exist in a given model, bound states of such solitons with an event horizon can typically be constructed (see, $e.g.$,~\cite{Kastor:1992qy}), leading to more complicated BHs, often called \textit{hairy}~\footnote{This is, however, no golden rule. A famous example are spherically symmetric boson stars, which admit no black hole generalization~\cite{Pena:1997cy}. This example emphasizes that symmetry compatibility between the soliton and horizon is a non-trivial necessary condition to place the latter inside the former.}. For instance, placing a horizon inside the two above examples of gravitating solitons, leads, respectively, to  Kerr BHs with scalar hair~\cite{Herdeiro:2014goa,Herdeiro:2015gia}  and ``coloured" BHs~\cite{Volkov:1989fi,Volkov:1990sva,1990JMP....31..928K,Bizon:1990sr,Volkov:1998cc}.

This general principle indicates how departures from Israel's theorem can be constructed, using the fact that solitonic objects allow, typically, less symmetries. Indeed, explicit  static gravitating solitons and BHs with \textit{only axial symmetry} 
were constructed $e.g.$ in \cite{Kleihaus:1996vi,Kleihaus:1997ic} within EYM theory. 
But something even more dramatic should be possible. A number of (non-linear) field theories possess, 
on a Minkowski background, known static solitonic solutions \textit{without any} continuous (spatial) symmetries~
($e.g.$~\cite{Houghton:1995bs,Faddeev:1996zj,Battye:1997qq,Battye:1998zn}), 
which must gravitate when coupled to GR. 
The addition of a horizon, therefore, will likely yield static BHs without any continuous (spatial) symmetries. 
Up to now, however, this maximal departure from Israel's theorem found no explicitly constructed realization;  see~\cite{Ridgway:1995ke,Ioannidou:2006mg} for partial results in this direction.

Recently, a new candidate model for this construction was unveiled: $AdS$-electrovacuum. 
Classical results in GR established the inexistence of  gravitating solitons in vacuum~\cite{Einstein:1943:NER,Lichnerowicz}, electrovacuum~\cite{Heusler:1996ft}, or $AdS$-vacuum~\cite{Boucher:1983cv}. Remarkably, in $AdS$-\textit{electrovacuum}, 
and despite apparent obstructions~\cite{Shiromizu:2012hb}, such solitons exist naturally. 
They were anticipated and constructed linearly in~\cite{Herdeiro:2015vaa} 
by simple considerations of electrostatics in global $AdS$; 
fully non-linearly examples were presented in~\cite{Costa:2015gol} and~\cite{Herdeiro:2016xnp}. 
In a nutshell: 
$(i)$ the box-like structure of $AdS$ allows the existence of electric (or magnetic) multipoles, 
as test fields, which are everywhere regular. 
They are defined by their multipole structure at the $AdS$ boundary. 
$(ii)$ Their backreaction yields Einstein-Maxwell-$AdS$ solitons, 
which inherit the spatial symmetries of the boundary data. 
$(iii)$ Introducing a horizon yields a static BH without continuous spatial symmetries, 
for appropriate boundary multipoles. 
A static, axially symmetric, 
BH within a dipole soliton was constructed in~\cite{Costa:2015gol}. 
Here, we construct static BHs without any spatial isometry, 
which, as we shall see, require solitons with higher multipoles.

\noindent{\bf{\em Smooth electric multipoles on $AdS$-electrovacuum.}}
Einstein-Maxwell theory with 
a negative cosmological constant  is described by the action:
\begin{eqnarray}
\mathcal{S} =\frac{1}{16\pi G} \int d^4 x\sqrt{-g}\left\{  R-2\Lambda 
 - F_{\mu \nu}F^{\mu\nu}\right\} \ .
 \label{EMAdS}
\end{eqnarray}
 $F=dA$ is the $U(1)$ field strength and $\Lambda\equiv -3/L^2<0$ is the cosmological constant, where $L$ is the $AdS$ ``radius". Varying the action one obtains the Einstein-Maxwell equations, $G_{\mu\nu}+\Lambda g_{\mu\nu}=2 T_{\mu\nu}$, $d\star F=0$, where the electromagnetic energy-momentum tensor is $T_{\mu\nu}=F_{\mu \alpha}F_{\nu\beta}g^{\alpha \beta}-g_{\mu\nu}F^2/4$. 
The maximally symmetric solution of this theory is $AdS$, with $F=0$,
 which in global coordinates reads
\begin{eqnarray}
\label{AdS}
ds^2=-N(r)dt^2+\frac{dr^2}{N(r)}+r^2(d \theta^2+\sin^2\theta d\varphi^2) \ ,
\end{eqnarray}  
where $N(r)=1 +{r^2}/{L^2}$.

$AdS$ electrostatics in global coordinates, for test fields, exhibits an important difference $w.r.t.$ its Minkowski counterpart: there are everywhere regular solutions for all multipoles (except the monopole), which decay  as $1/r$, asymptotically~\cite{Herdeiro:2015vaa}. A similar statement applies to magnetostatics~\cite{Herdeiro:2016xnp}. In these previous studies, 
only the axi-symmetric multipoles
were considered. 
Here we consider the most general electrostatic potential,  $A= V(r,\theta,\varphi)dt$, with: 
\begin{equation}
 V(r,\theta,\phi)= \sum_{\ell\geqslant 1}\sum_{m=-\ell}^{m=\ell}c_{\ell m}R_\ell(r)Y_{\ell m}(\theta,\varphi) \ ,
\end{equation}
where $c_{\ell m}$ are
arbitrary constants and $Y_{\ell m}(\theta,\varphi)$ are the $real$ spherical harmonics
\cite{Whittaker},
normalized such that
 $ \int d\Omega~Y_{\ell m} Y_{\ell' m'} =\delta_{\ell \ell'}\delta_{mm'}$.
Due to the $AdS$ background symmetries, the radial equation is $m$-independent,
$
 \frac{d}{dr}\left(r^2\frac{d R_\ell(r)}{dr}\right)=\frac{1}{N(r)}\ell(\ell+1)R_{\ell}(r) .
$
For $\ell\geqslant 1$ this equation possesses a solution 
which is regular everywhere (in particular at $r=0$), that can be written in terms of hypergeometric functions~\cite{Herdeiro:2015vaa}:
\begin{eqnarray}
R_\ell(r)=
\frac{\Gamma(\frac{1+\ell}{2})\Gamma(\frac{3+\ell}{2})}{\sqrt{\pi}\Gamma\left(\frac{3}{2}+\ell\right)}
 \frac{r^{\ell}}{L^\ell}~{}_2F_1\left(\frac{1+\ell}{2}, \frac{\ell}{2},\frac{3}{2}+\ell,- \frac{r^2}{L^2}\right) \ , \nonumber
\end{eqnarray} 
where the normalization guarantees that $R_\ell(r)\to 1$ asymptotically.

The energy density of the solutions, $\rho=-T^t_t$, is finite everywhere
and strongly localized in a finite region of space,
depending on $both$ $\theta$ and $\varphi$.
 $\rho$ is nonzero at $\theta=0$;
at $r=0$ it vanishes unless $\ell=1$.
At infinity, $\rho$ decays as $1/r^4$, such that
 the   total energy of these solutions, $E=-\int \sqrt{-g}T_t^t d^3 x$, is finite.  
With the chosen normalization $
E_\ell=L{ \Gamma(\frac{1+\ell}{2}) \Gamma(\frac{3+\ell}{2})}/[ \Gamma(1+\frac{\ell}{2}) \Gamma(\frac{\ell}{2})]$~\footnote{This value differs from the one in~\cite{Herdeiro:2015vaa} due to a different normalization of the spherical harmonics.}.

These static   
regular electric multipoles on a fixed $AdS$ background satisfy the virial identity
   \begin{eqnarray}
 \int_0^\infty r^2 dr \int_0^\pi \sin \theta
 \left[
 V_{,r}^2+\frac{1-\frac{r^2}{L^2}}{N^2(r)r^2}\left( V_{,\theta}^2+\frac{ V_{,\varphi}^2}{\sin^2\theta}\right)
\right]=0 \ .
\nonumber
\end{eqnarray}
In the $L\rightarrow \infty$ limit (Minkowski), all terms in the integrand are positive definite and no non-trivial configurations can exist. This identity clarifies that: $(i)$  the $AdS$ geometry supplies the attractive force needed to balance the repulsive gauge interactions; $(ii)$ the configurations are supported by the nontrivial angular dependence of $V$, $i.e.$ they must possess a multipolar structure.

\begin{figure}[h!]
\begin{center}
\includegraphics[width=0.09\textwidth]{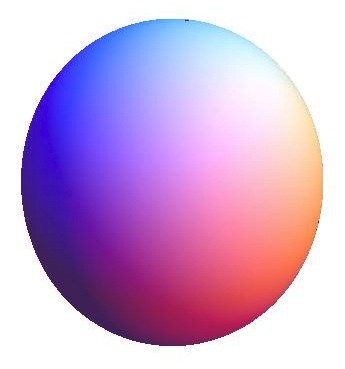} 
\includegraphics[width=0.09\textwidth]{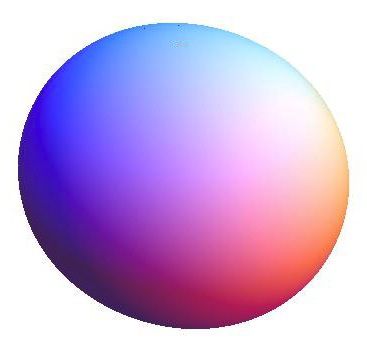} 
\\
\includegraphics[width=0.07\textwidth]{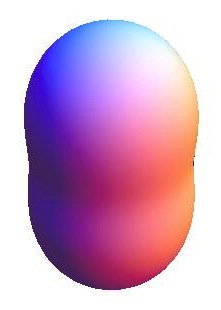} 
\includegraphics[width=0.10\textwidth]{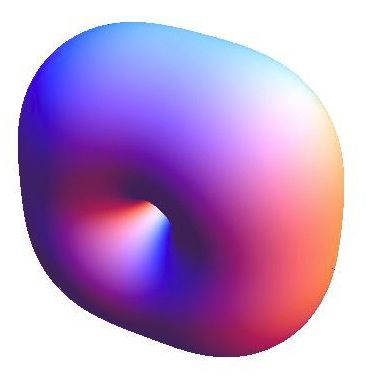} 
\includegraphics[width=0.10\textwidth]{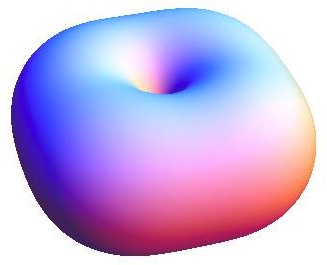} 
\\
\includegraphics[width=0.07\textwidth]{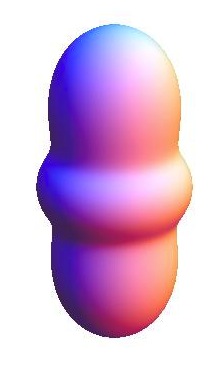} 
\includegraphics[width=0.09\textwidth]{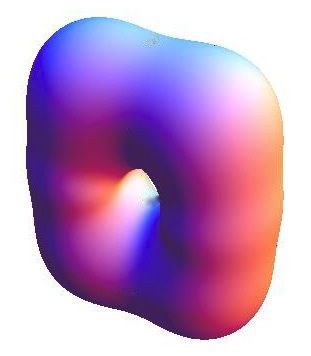}
\includegraphics[width=0.09\textwidth]{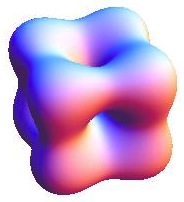}\ \ 
\includegraphics[width=0.10\textwidth]{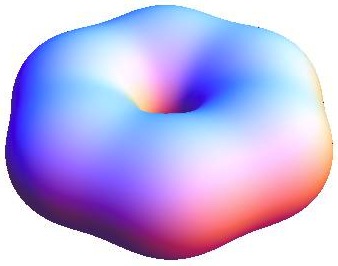}
\\
\includegraphics[width=0.06\textwidth]{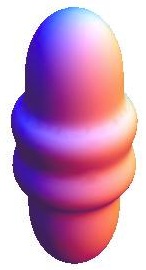} 
\includegraphics[width=0.085\textwidth]{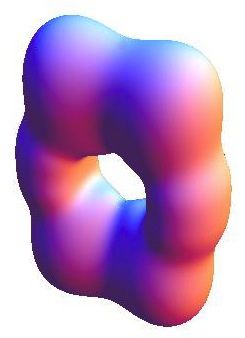} 
\includegraphics[width=0.085\textwidth]{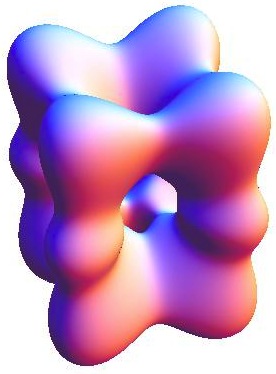} \
\includegraphics[width=0.105\textwidth]{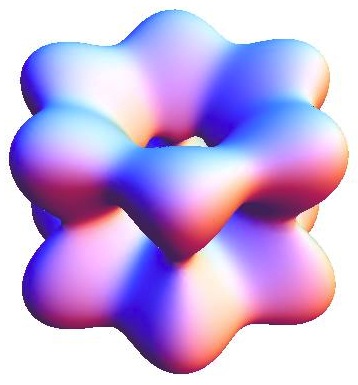} 
\includegraphics[width=0.12\textwidth]{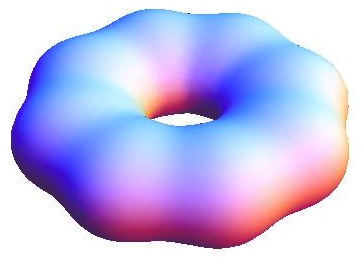} 
\caption{Examples of surfaces of constant energy density for the Maxwell-$AdS$ regular electric multipoles with  (from left to right): 
$(\ell,m)=\left\{(1,0), (1,1)\right\}$ (top row); 
$(\ell,m)=\left\{(2,0), (2,1), (2,2)\right\}$ (second row);  
$(\ell,m)=\left\{(3,0), (3,1), (3,2), (3,3)\right\}$ (third row); 
$(\ell,m)=\left\{(4,0), (4,1), (4,2), (4,3), (4,4)\right\}$ (fourth row). 
Here, we are defining standard Cartesian coordinates from the global $AdS$ coordinates, using the standard formulas.
All plots in this work use units with $L=G=1$.}
\label{figure_probe}
\end{center}
\end{figure}

In Fig.~\ref{figure_probe} we exhibit surfaces of constant energy density  for a sample of these solutions, with $\ell=1,2,3,4$
and $m\geqslant 0$
(the case $m<0$ follows directly).
 For $m\neq 0$, these surfaces possess solely discrete symmetries. 
The exception to this pattern occurs for $\ell=1$, 
wherein the $m=1$ and $m=0$ multipoles are related by a rotation 
(as indeed are the $\ell=2$, $m=1$ and $m=2$ multipoles). 
Thus, obtaining static BHs with no spatial isometries requires taking $\ell \geqslant 2$.

Similar solutions to the ones just described are found when taking instead a Schwarzschild-$AdS$ ($SAdS$) BH backgound,  with a line element still given by 
(\ref{AdS}) where now 
$N(r)=(1-{r_H}/{r})[1+(r^2+r r_H+r_H^2)/L^2]$, and  $r_H>0$ is the event horizon radius. The corresponding radial equation 
cannot, however, be solved in closed form any longer (except for $\ell=0$). 
But it is straightforward to obtain a numerical solution, for any $\ell\geqslant 1$~\cite{Kichakova:2015nni}. 
The radial function vanishes on the horizon, in the neighbourhood of which it 
can be written as a power series in $(r-r_H)$. 
Solutions are regular everywhere, on and outside the horizon, 
showing that these regular electric multipoles can be superimposed on the $SAdS$ BH.

\noindent{\bf{\em The non-linear setup.}}
%
Fully non-linear $AdS$-electrovacuum solitons and BHs are obtained from the backreaction of the above solutions. We shall tackle the fully non-linear Einstein-Maxwell-$AdS$ equations numerically, employing the Einstein-De Turck (EDT) method~\cite{Headrick:2009pv,Adam:2011dn}. This approach to the numerical treatment of stationary problems in GR does not require fixing, \textit{a priori}, a metric gauge, 
yielding, nevertheless, elliptic equations (see, $e.g.$~\cite{Wiseman:2011by,Dias:2015nua} for reviews).   The EDT equations are:
\begin{eqnarray}
\label{EDT}
R_{\mu\nu}-\nabla_{(\mu}\xi_{\nu)}=\Lambda g_{\mu\nu}+2 \left(T_{\mu\nu}-\frac{1}{2}T  g_{\mu\nu}\right) \ .
\end{eqnarray}
Here, $\xi^\mu$ is a vector defined as
$
\xi^\mu\equiv g^{\nu\rho}(\Gamma_{\nu\rho}^\mu-\bar \Gamma_{\nu\rho}^\mu),
$
where 
$\Gamma_{\nu\rho}^\mu$ ($\bar \Gamma_{\nu\rho}^\mu$) is the Levi-Civita connection associated to the
spacetime metric $g$ that one wants to determine (a reference metric $\bar g$ that is introduced).
Solutions to (\ref{EDT}) solve the Einstein equations
iff $\xi^\mu \equiv 0$ everywhere on the manifold.

To solve~\eqref{EDT}, together with the Maxwell equations, we use  an  ansatz
with seven unknown metric functions, $F_1,F_2,F_3,F_0,S_1,S_2,S_3$
and an electrostatic potential $V$:
\begin{eqnarray}
\nonumber
&&
ds^2=F_1(r,\theta,\varphi) \frac{dr^2}{N(r)}+F_2(r,\theta,\varphi)\left[r d\theta+S_1(r,\theta,\varphi) dr \right]^2 
\\
\nonumber
&&
+F_3(r,\theta,\varphi)  \big[r \sin \theta d\varphi+S_2(r,\theta,\varphi) dr+S_3(r,\theta,\varphi) r d\theta \big]^2
\\
&&
-F_0(r,\theta,\varphi) N(r)dt^2 \ ,~~~{\rm and}~~~A=V(r,\theta,\varphi) dt\ ,
 \label{metric}
\end{eqnarray}
where $N(r)=\left(1-\frac{r_H}{r}\right)\left(1+\frac{r^2+r r_H+r_H^2}{L^2}-\frac{q^2}{r r_H}\right)$ is a background function, 
with $r_H> 0$ the event horizon radius and $q$ another input constant.
Then the problem reduces to
solving a set of eight PDEs with suitable
boundary conditions (BCs).
The BCs are found by 
 constructing an approximate form of the solutions on the
boundary of the domain of integration, compatible with the requirement $\xi^\mu = 0$, 
plus
regularity and AdS asymptotics.
In particular, the first requirement should imply $\xi^\mu \equiv 0$ \textit{everywhere},
a condition which is verified from  the numerical output. 

We have focused our study on $m>0$ solutions with a reflection symmetry along the equatorial plane
($\theta=\pi/2$)
and two $\mathbb{Z}_2$-symmetries $w.r.t.$ the $\varphi-$coordinate.
The domain of integration  for the $(\theta,\varphi)$-coordinates 
is then $[0,\pi/2]\times [0,\pi/2]$. Explicitly, we impose the following
 BCs at infinity 
$F_0=F_1=F_2=F_3=1$,
$S_1=S_2= S_3=0,~V=c_e Y_{\ell m}(\theta,\varphi)$, 
which defines the Maxwell boundary data to be a single harmonic, $(\ell,m)$, 
with strength $c_e$. 
The BCs at $\theta=0$ are
$\partial_\theta F_0=\partial_\theta F_1= \partial_\theta F_2=  \partial_\theta F_3=0,$
$S_1=S_2= \partial_\theta S_3=0,~V=0$.
At $\theta=\pi/2$ we impose
$\partial_\theta F_0=\partial_\theta F_1= \partial_\theta F_2=  \partial_\theta F_3= 0,$
$S_1=\partial_\theta S_2= S_3=0$, together with $ V=0$,
except if  $\ell+m$ is an even number, in which case we impose  
$\partial_\theta  V=0$.
The BCs at $\varphi=0$ are
$\partial_\varphi F_1= \partial_\varphi F_2=  \partial_\varphi F_3= \partial_\varphi F_0=0,~
\partial_\varphi S_1=   S_2=   S_3=0,~\partial_\varphi  V=0$.
At
 $\varphi=\pi/2$ 
we impose
$\partial_\varphi F_1= \partial_\varphi F_2=  \partial_\varphi F_3= \partial_\varphi F_0=0,~
\partial_\varphi S_1=   S_2=   S_3=0$
 together with $ V=0$ for odd $m$, or  $\partial_\varphi  V=0$ for even $m$.
Solitonic solutions have $r_H=0=q$ and 
the range of the radial coordinate is $0\leqslant r<\infty$.
At $r=0$ we impose
$\partial_r F_1= \partial_r F_2=  \partial_r F_3= \partial_r F_0= 
\partial_r S_1=\partial_r S_2= \partial_r S_3=0,~V=0.$
The BHs have a horizon located at  
$r= r_H > 0$.
To deal with the  BCs there, 
it proves useful to introduce a new (compact) radial
coordinate  $x$, as $r\equiv \frac{r_H}{1-(\frac{x}{2L})^2}$, such that $0\leqslant x<2L$ and in terms of which the horizon is located at $x=0$.
This yields the following BCs at the horizon: 
$\partial_x F_1=\partial_x F_2=\partial_x F_3=\partial_x F_0=0,~
S_1=S_2=\partial_x S_3=0,~V=0.$

\noindent{\bf{\em Numerical Procedure.}}
We have successfully obtained numerical solutions for both   BHs and solitons in $AdS$-electrovacuum, 
fixing the gauge field boundary data to be a single $Y_{\ell m}$ harmonic, 
and scanning through a variety of $\ell,m$ values.  
The numerical procedure we have used is 
a modified version of the approach previously employed
in the study of axially symmetric configurations
of the same model~\cite{Herdeiro:2016xnp}. 
The field equations are first discretized on a $(r, \theta,\varphi)$ grid with 
$N_r\times N_\theta \times N_\varphi$
points.
The grid spacing in the $r$-direction is non-uniform, whilst the values of the grid points in the angular
directions are uniform.
Typical grids have sizes 
$\sim 100 \times 30 \times 30$.
The resulting system is solved
iteratively until convergence is achieved. 
Computations are performed by adapting a finite difference code described in~\cite{schoen}
based on the iterative Newton-Raphson
method. 
For the solutions herein,
 the typical numerical error is estimated to be
$\lesssim 10^{-3}$.

In practice, the BH solutions are found starting with 
$SAdS$ BHs and slowly increasing the 
parameter $c_e$ in the BCs at infinity. 
In a second step, the parameters $(r_H,q)$ in (\ref{metric})
are also varied.

\noindent{\bf{\em Horizon geometry.}}
The most unusual property of the generic BH solutions
is that their horizons do not possess a rotational symmetry,
despite being topologically a 2--sphere. 
To establish this result, we consider the
induced metric at the  horizon, which reads, from~\eqref{metric}, 
\begin{eqnarray}
&
d\sigma^2=r_H^2 \left[
 F_2 d\theta^2+ F_3  (\sin \theta d\varphi+  S_3  d\theta)^2
\right] , 
\label{horizonm}
  \end{eqnarray}
 where $F_2,F_3,S_3$ are now only functions of $\theta,\varphi$.
To visualize  this geometry, we consider
its isometric embedding in a flat three-dimensional space,
with $d\sigma^2=dX^2+dY^2+dZ^2$, 
the embedding functions,   
$X(\theta,\varphi)$,
$Y(\theta,\varphi)$,
$Z(\theta,\varphi)$, 
being found by integrating a
system of non-linear PDEs.
 \begin{figure}[h!]
\begin{center}
\includegraphics[width=0.16\textwidth]{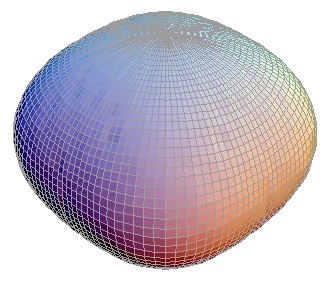}
\includegraphics[width=0.15\textwidth]{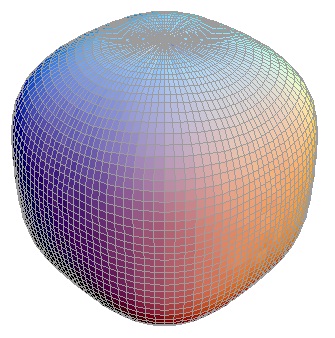} 
\includegraphics[width=0.16\textwidth]{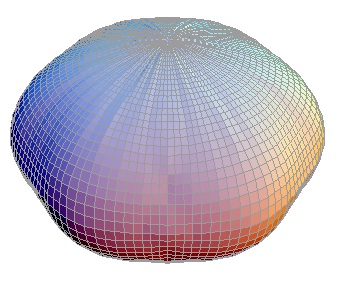} 
\\ 
\vspace{0.1cm}
\hspace{-0.4cm}
\includegraphics[width=0.12\textwidth]{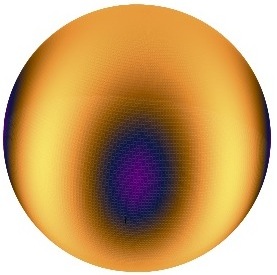} {~~~~~~}
\includegraphics[width=0.12\textwidth]{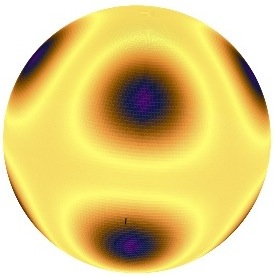} {~~~~~~}
\includegraphics[width=0.12\textwidth]{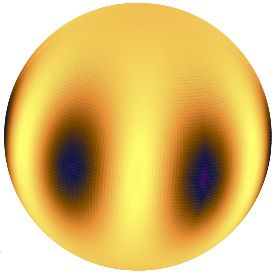}
\caption{Examples of isometric embeddings for the horizon of $AdS$-electrovacuum BHs (top), together with their horizon Ricci scalar 
(bottom). 
The boundary data is given by the harmonics with $(\ell,m)$ equal to $(2,2)$ (left),  $(3,2)$ (middle) 
and  $(3,3)$ (right). }
\label{figure_nonlinear_na}
\end{center}
\end{figure}
%
\begin{figure}[h!]
\begin{center}
\includegraphics[width=0.235\textwidth]{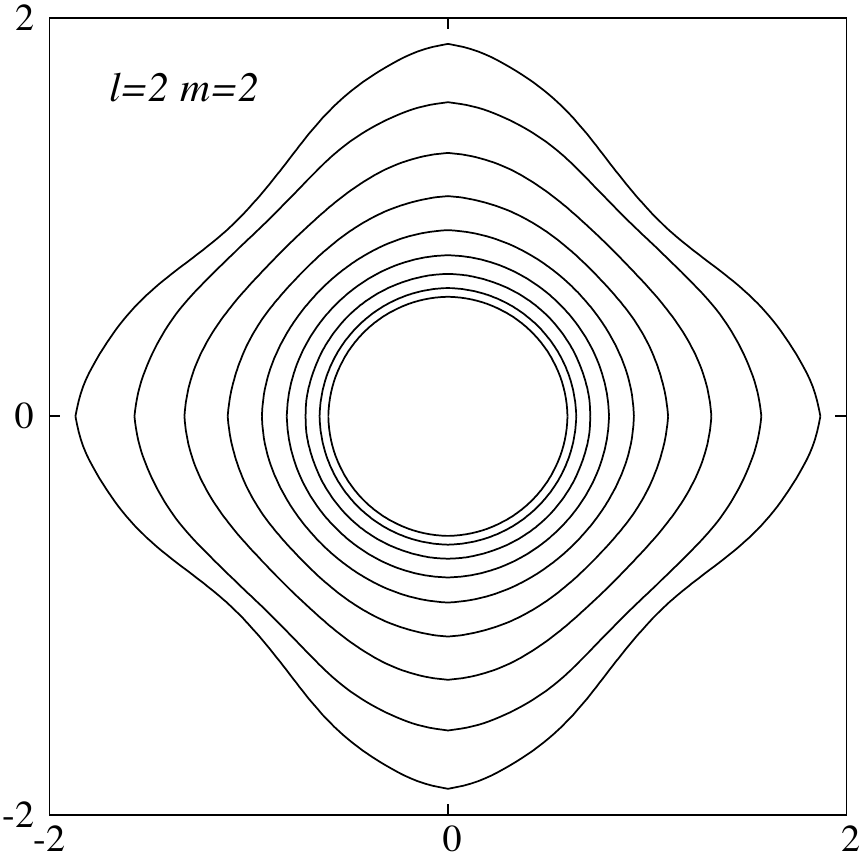} 
\includegraphics[width=0.235\textwidth]{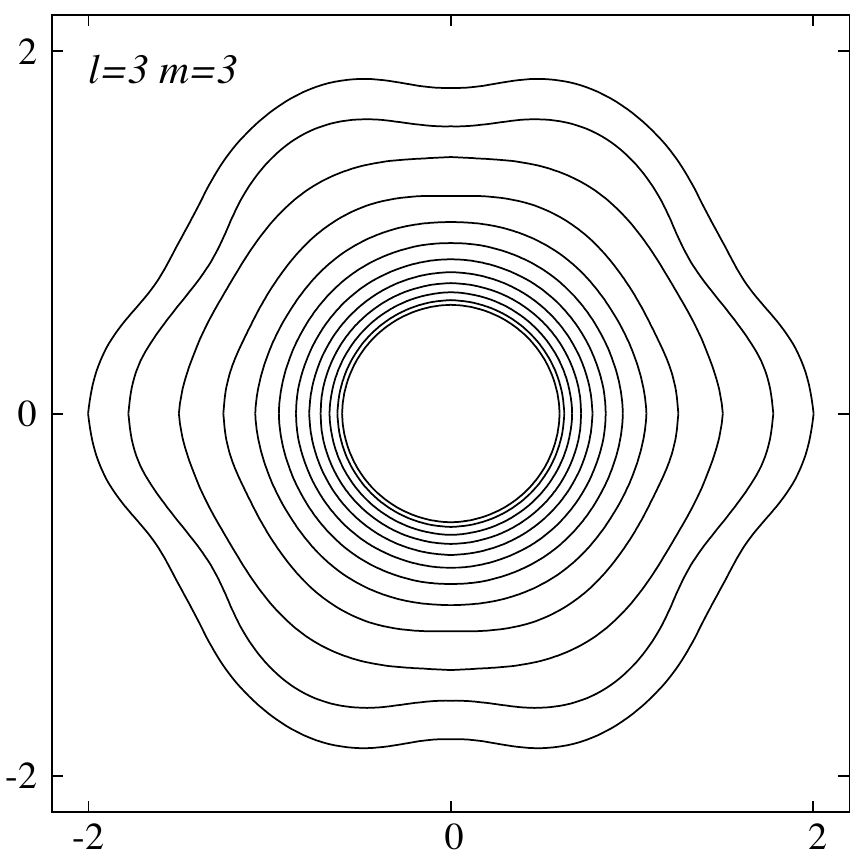} 
 \caption{
Equatorial slices for isometric embeddings of the horizons of $AdS$-electrovacuum BHs 
with different boundary data.
The BHs have the same temperature 
and increasing values of the parameter $c_e$,
starting with $c_e=0$ (center).
}
 \label{figure_slice}
 \end{center}
 \end{figure}
%
%
In Fig.~\ref{figure_nonlinear_na} we exhibit 3D isometric 
embeddings for a set of non-axisymmetric BHs. 
For instance, for $(\ell,m)=(3,2)$ boundary data, one obtains a \textit{cubic-like} horizon. 
Comparing with Fig.~\ref{figure_probe} one observes that the horizon loosely adapts 
to the corresponding constant energy surface, except that it is topologically simply connected. 
Also,
the horizon scalar curvature is everywhere finite, although it can take large values~\footnote{As a further test of our solutions, we have verified the Gauss-Bonnet theorem holds at the level of the numerical accuracy, by integrating the horizon Ricci scalar over the horizon.
}.
As shown in Fig. \ref{figure_slice}, the horizon deformation increases with $c_e$;
these global isometric embeddings, however, can only be obtained up to some threshold value of $c_e$, 
beyond which well known obstructions arise (see $e.g.$~\cite{Smarr:1973zz,Gibbons:2009qe}).

\noindent{\bf{\em Global charges and thermodynamics.}}
These  configurations carry a nonzero 
electric charge density;  their total electric charge, however, vanishes.
As such, 
the only global charge of the solutions is their mass $M$.
Its expression, computed by employing either the prescriptions  in  
\cite{Balasubramanian:1999re} or the one in 
\cite{Ashtekar:1999jx},  
is
\begin{eqnarray}
\nonumber
 M=M^{(b)}-\frac{3L}{16\pi G} 
\int_0^{2\pi} d\varphi  \int_0^\pi d\theta  \sin \theta
f_{03}(\theta,\varphi)\ ,
 \end{eqnarray}
where
$M^{(b)}=\frac{r_H }{2G}\left(1+\frac{r_{H}^2}{L^2}+\frac{q^2}{r_H^2} \right)$
is a contribution from the background metric and $f_{03}(\theta,\varphi)$
is a function which enters the far field asymptotics, with $F_0=1+f_{03}/r^3+\dots$. 
%
\begin{figure}[h!]
\begin{center}
\includegraphics[width=0.5\textwidth]{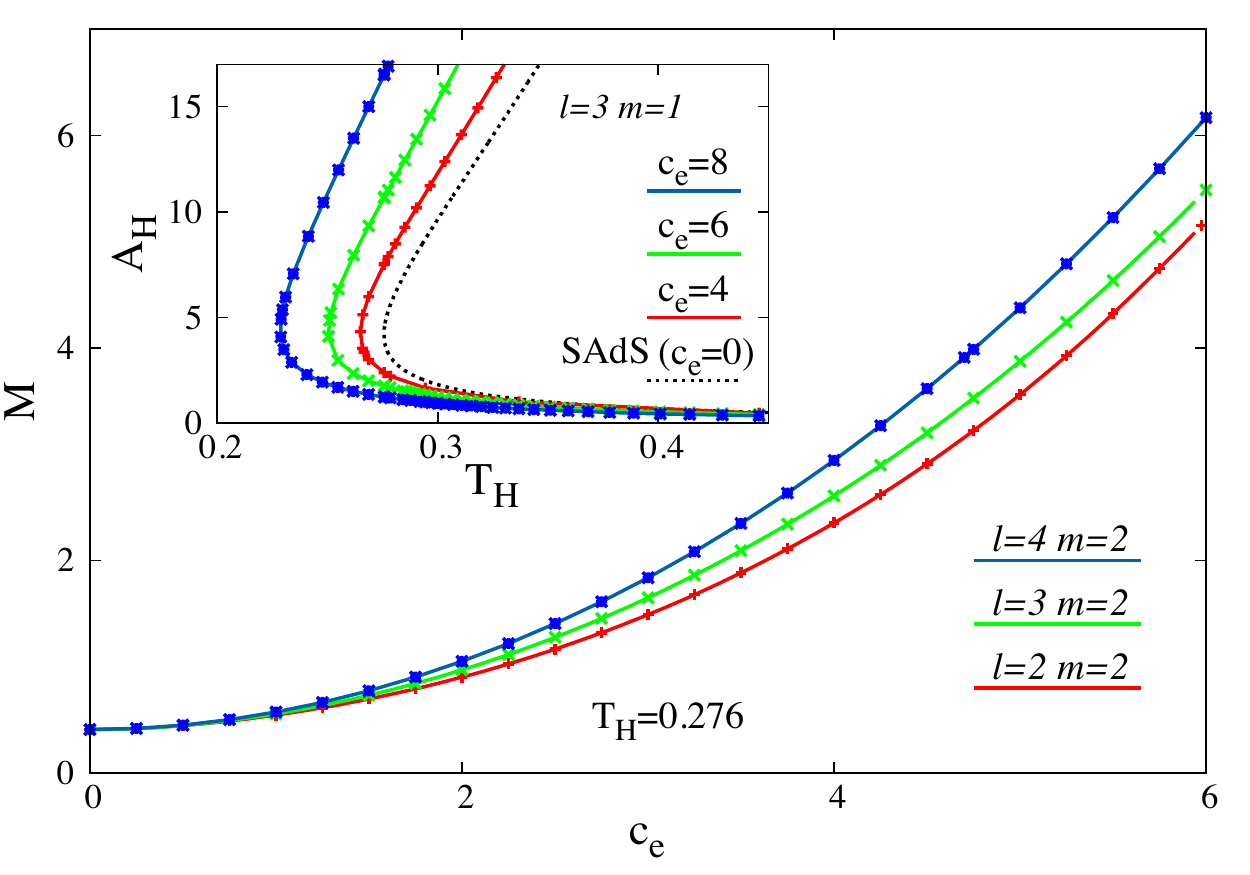}\ \ \
\caption{
Mass $vs.$ $c_e$  for families of BHs
with different boundary data and the same temperature. 
(Inset) Horizon area $vs.$ temperature 
for $(3,1)$ BHs with different values of $c_e$.
}
\label{figure_nonlinear_energy}
\end{center}
\end{figure}
%
In Fig.~\ref{figure_nonlinear_energy} we exhibit 
the total mass for BH solutions with different $(\ell,m)$ values of boundary data, 
for a fixed temperature and varying $c_e$. 
The pattern is universal: the mass increases with $c_e$
and also (for the same $m$) with the multipole number ${\ell}$.

Of interest are also the
horizon area and Hawking temperature of the BHs,  
 \begin{eqnarray}
\nonumber
 A_H= \int_0^{2\pi} d\varphi \int_0^{\pi} d\theta \sqrt{g_\sigma} \ , \ \ \ \ 
 T_H= \frac{ (1+ {3r_H^2}/{L^2}-q^2 )}{4\pi r_H } \ ,
 \end{eqnarray} 
where $\sqrt{g_\sigma}=r_H^2 \sin\theta \sqrt{F_2  F_3}$.
In the absence of a net electric charge, the thermodynamics has similarities to that of $SAdS$ BHs.
As shown in Fig.~\ref{figure_nonlinear_energy} (inset),  
 there are two branches of BHs,
 existing above a minimal temperature $T_H^{(min)}>0$,   
where $T_H^{(min)}$ decreases with $c_e$.
For lower branch solutions,  
the BH size decreases with $T_H$,
while for upper branch BHs, the horizon area increases with the temperature, 
with no upper bound on $A_H$.

These BH solutions possess a 
nontrivial zero horizon size limit $r_H\to 0$,
 corresponding to {\it AdS-electrovacuum solitons
with no isometries}.
The $(M,c_e)$-diagram of the solitons is similar to that
exhibited for BHs in Fig.~\ref{figure_nonlinear_energy}.

Finally, let us mention two generalizations: $(i)$ 
these BHs can be endowed with a net electric charge
by turning on an additional ${\ell}=0$ mode, in the 
boundary condition for $V$ at infinity.
Such solutions, however, do not possess a solitonic limit
and can be thought of as
 describing the (nonlinear) superposition
of a Reissner-Nordstr\"om (rather than Schwarzschild) BH with $AdS$-electrovacuum solitons; $(ii)$ the configurations described herein will possess a magnetic dualized version, which can be straighforwardly constructed.

\noindent{\bf{\em Remarks.}}
Static (single) BHs in electrovacuum can only have an electric monopole, and they are necessarily spherically symmetric. In sharp contrast, static BHs in $AdS$-electrovacuum can have an \textit{arbitrary electric multipole structure}; by turning on appropriate multipoles, we have presented explicit examples of static BHs with no continuous (spatial) symmetries. 

The BHs presented here still exhibit discrete symmetries. Is it possible, with appropriate boundary data, to obtain BH horizons, in $AdS$-electrovacuum, isometric to any topologically spherical 2-manifold? If not, what 2-geometries are allowed? Whatever the correct answer is, the results reported herein show (yet) another example of how conceptually different $AdS$ gravity is from its Minkowski space counterpart.

\bigskip

\noindent{\bf{\em Acknowledgements.}} The authors acknowledge funding from the FCT-IF programme. This work was partially supported by  the  H2020-MSCA-RISE-2015 Grant No.  StronGrHEP-690904, and by the CIDMA project UID/MAT/04106/2013. Computations were performed at the Blafis cluster, in Aveiro University.

\bibliographystyle{h-physrev4}
\bibliography{letter}
\end{document}